\documentclass{ws-procs975x65}

\begin{document}

\title{Hubble Meets Planck: A Cosmic Peek at Quantum Foam}

\author{Y. Jack Ng}
\address{Institute of Field Physics, Department of Physics and Astronomy,\\
University of North Carolina, Chapel Hill, NC 27599-3255, USA\\
E-mail: yjng@physics.unc.edu}

\begin{abstract}

If spacetime undergoes quantum fluctuations, an electromagnetic wavefront
will acquire uncertainties
in direction as well as phase as it propagates through spacetime.
These uncertainties can show up in interferometric observations of distant
quasars as a decreased fringe visibility.  The Very
Large Telescope and Keck interferometers may be on the verge of probing
spacetime fluctuations which, we also argue, have repercussions for
cosmology, requiring the existence of dark energy/matter, the critical cosmic
energy density, and accelerating cosmic expansion in the present era.

\end{abstract}


\bodymatter

\section{Quantum Fluctuations of Spacetime}

Conceivably spacetime, like everything else,
is subject to quantum fluctuations.  As a result, spacetime is ``foamy" at
small scales,
\cite{Wheeler} giving rise to a microscopic structure of spacetime known as
quantum foam, also known as spacetime foam, and
entailing an intrinsic limitation $\delta l$ to the accuracy
with which one can measure a distance $l$.  In principle, $\delta l$ can
depend on both
$l$ and the Planck length $l_P = \sqrt{\hbar G/c^3} $, the intrinsic scale
in
quantum gravity,
and hence can be written as
$\delta l \gtrsim l^{1 - \alpha} l_P^{\alpha}$, with $\alpha
\sim 1$ parametrizing the various spacetime foam models.  (For related
effects
of quantum fluctuations of spacetime geometry, see Ref. 2.)
In what follows, we will advocate the so-called
holographic model corresponding to $\alpha = 2/3$, but we will also consider
the (random walk) model with $\alpha = 1/2$ for comparison.
The holographic model has been derived by various arguments,
including the Wigner-Saleckar gedankan experiment to measure a distance
\cite{wigner} and the
holographic principle \cite{ng,stfoam}. (See my contribution to the
Proceedings of MG10 \cite{MG10}.)
Here in the two subsections to follow, we use instead (1) an approach based
on quantum computation, and (2) an argument over the maximum
number of particles that can be put inside a region of space
respectively.

\subsection{Quantum Computation}

This method\cite{llo04,gio04}
hinges on the fact that quantum fluctuations of spacetime
manifest themselves in the form of uncertainties in the geometry of
spacetime.  Hence the structure of spacetime foam can be inferred from the
accuracy with which we can measure that geometry.  Let us
consider a spherical volume of
radius $l$ over the amount of time $T = 2l/c$ it takes light to cross the
volume.  One way to map out the geometry of this spacetime region is to fill
the space with clocks, exchanging
signals with other clocks and measuring the signals' times of arrival.
This process of mapping the geometry is a sort of computation; hence
the total number of operations (the ticking of the clocks and
the measurement of signals etc) is bounded by the Margolus-Levitin
theorem\cite{Lloyd}
in quantum computation,
which stipulates that the rate of operations for any computer
cannot exceed the amount of energy $E$ that is available for computation
divided by $\pi \hbar/2$.   A total mass $M$ of clocks then
yields, via the Margolus-Levitin theorem, the bound on the total number of
operations given by $(2 M c^2 / \pi \hbar) \times 2 l/c$.  But to prevent
black hole formation, $M$ must be less than $l c^2 /2 G$.  Together, these
two limits imply that the total number of operations that can occur in a
spatial volume of radius $l$ for a time period $2 l/c$ is no greater than
$\sim (l/l_P)^2 $.  (Here and henceforth we neglect multiplicative constants
of order unity, and set $c=1=\hbar$.)
To maximize spatial resolution, each clock must tick
only once during the entire time period.  And if we regard the operations
partitioning the spacetime volume into ``cells", then on the average each
cell
occupies a spatial volume no less than $ \sim l^3 / ( l^2 / l_P^2) =
l l_P^2 $, yielding an average separation between neighhoring cells
no less than $l^{1/3} l_P^{2/3}$.  This spatial separation
is interpreted as the {\it average minimum uncertainty} in the measurement
of
a distance $l$, that is, $\delta l \gtrsim l^{1/3} l_P^{2/3}$.

Parenthetically we can now understand why this quantum foam model has come
to be known
as the holographic model.
Since, on the average, each cell occupies a spatial volume of $l l_P^2$,
a spatial region of size $l$ can contain no more than $l^3/(l l_P^2) =
(l/l_P)^2$ cells.  Thus this model
corresponds to the case of
maximum number of bits of information $l^2 /l_P^2$
in a spatial region of size $l$, that is
allowed by the holographic principle \cite{wbhts}, acording to which,
the maximum amount of information stored in a region of space scales as
the area of its two-dimensional surface, like a hologram.

It will prove to be useful to compare the holographic model in the mapping
of the geometry of spacetime
with the one that corresponds to spreading the spacetime cells uniformly
in both space and time.  For the latter case, each cell has
the size of $(l^2 l_P^2)^{1/4} =
l^{1/2} l_P^{1/2}$ both spatially and temporally, i.e., each clock ticks
once in the time it takes to communicate with a neighboring clock.  Since
the dependence on $l^{1/2}$ is the hallmark of a random-walk fluctuation,
this quantum foam model corresponding to  $\delta l \gtrsim
(l l_P)^{1/2}$ is called the random-walk model \cite{AC}.
Compared to the holographic model, the random-walk model predicts a
coarser spatial resolution, i.e., a larger distance fluctuation,
in the mapping of spacetime geometry.  It
also yields a smaller bound on the information content in a spatial
region, viz., $(l/l_p)^2 / (l/l_P)^{1/2} = (l^2 / l_P^2)^{3/4} =
(l/l_P)^{3/2}$.

\subsection{Maximum Number of Particles in a Region of Space}

This method involves an estimate of the maximum number of particles that can
be
put inside a spherical region of radius $l$.  Since matter can embody
the maximum information when it is converted to energetic and effectively
massless particles, let us consider massless particles.
According to Heisenberg's uncertainty
principle, the minimum energy of each particle is no less than
$\sim l^{-1}$.  To
prevent the region from collapsing into a black hole, the total energy is
bounded by $ \sim l/G$.
Thus the total number of particles must be less than $ (l/l_P)^2$, and
hence the average interparticle distance is no less than
$\sim l^{1/3} l_P^{2/3}$.  Now, the more particles
there are (i.e., the shorter the interparticle distance), the more
information can
be contained in the region, and accordingly the more accurate the geometry
of the region can be mapped out.  Therefore
the spatial separation we have just found can be interpreted as the average
minimum uncertainty in the measurement of a distance $l$; i.e., $\delta l
\gtrsim l^{1/3} l_P^{2/3}$.

Two remarks are in order.  First,
this minimum $\delta l$ just found corresponds
to the case of maximum energy density $\rho \sim (l l_P)^{-2}$ for the
region not
to collapse into a black hole, i.e., the holographic model, in contrast to
the random-walk model and other models, requires, for its
consistency, the {\it critical energy density} which, in the cosmological
setting, is $(H/l_P)^2$ with $H$ being the Hubble parameter. Secondly,
the numercial factor in
$\delta l$, according to the four different methods alluded to above, 
can be shown to be
between 1 and 2, i.e., $\delta l \gtrsim l^{1/3} l_P^{2/3}$ to $ 2 l^{1/3}
l_P^{2/3}$.

\section{Probing Quantum Foam with Extragalactic Sources}

The Planck length $l_P \sim 10^{-33}$ cm is so short that we need an
astronomical (even cosmological) distance $l$ for its fluctuation $\delta
l$ to be detectable.
Let us consider light (with wavelength $\lambda$)
from distant quasars or bright active galactic nuclei. \cite{lie03,NCvD}
Due to the quantum fluctuations of spacetime, the wavefront, while planar,
is itself ``foamy", having random fluctuations in phase \cite{NCvD} $\Delta
\phi \sim 2 \pi \delta l / \lambda$ as well as
the direction of the wave vector \cite{CNvD} given by $\Delta \phi / 2 \pi
$.
\footnote{Using $k = 2 \pi/ \lambda  $, one finds that, over one
wavelength, the wave vector fluctuates by $\delta k = 2 \pi \delta \lambda /
\lambda^2 = k \delta
\lambda / \lambda   $. Due to space isotropy of quantum fluctuations, the
transverse and longitudinal components of the wave vector fluctuate by
comparable amounts. Thus, over distance $l$,
the direction of the wave vector fluctuates by
$\Delta k_T /k = \Sigma \delta \lambda / \lambda \sim \delta l / \lambda$.}
In effect, spacetime foam creates a
``seeing disk" whose angular diameter is $\sim \Delta \phi /2 \pi  $.  For
an interferometer with baseline length $D$, this means that dispersion will
be seen as a spread in
the angular size of a distant point source, causing a reduction in the
fringe
visibility when $\Delta \phi / 2 \pi \sim \lambda / D  $. For a quasar of 1
Gpc away, at infrared wavelength, the
holographic model predicts a phase fluctuation $\Delta \phi \sim 2 \pi
\times
10^{-9}  $ radians.  On the other hand, an infrared interferometer (like the
Very Large Telescope Interferometer) with $D \sim 100$ meters has $\lambda /
D \sim 5 \times 10^{-9}  $.
Thus, in principle, this method will allow the use of interferometry fringe
patterns to test the holographic model!  Furthermore, these tests can be
carried out without guaranteed time using archived high resolution, deep
imaging data on quasars, and possibly, supernovae from existing and
upcoming telescopes.

The key issue here is the
sensitivity of the interferometer.  The lack of observed fringes may simply
be due to the lack of sufficient flux (or even just
effects originated from the turbulence of 
the Earth's atmosphere) 
rather than the possibility that the
instrument has resolved a spacetime foam generated halo.
But, given sufficient sensitivity, the
VLTI, for example, with its
maximum baseline, presumably has sufficient resolution to detect spacetime
foam halos for low redshift quasars, and in principle, it can be even more
effective for the higher redshift quasars. Note that the test is simply a
question
of the detection or non-detection of fringes.  It is not a question of
mapping the structure of the predicted halo.

\section{From Quantum Foam to Cosmology}

In the meantime, we can use existing archived data on
quasars or active galactic nuclei
from the Hubble Space Telescope to test the quantum foam models. \cite{CNvD}
Consider the case
of PKS1413+135 \cite{per02}, an AGN for which the redshift is $z = 0.2467$.
With $l \approx 1.2$ Gpc and $\lambda = 1.6 \mu$m,
we \cite{NCvD} find $\Delta \phi \sim 10 \times 2 \pi$ and
$10^{-9} \times 2 \pi$ for the random-walk model and
the holographic model of spacetime foam respectively.
With $D = 2.4$ m for HST, we expect to detect halos
if $\Delta \phi \sim 10^{-6} \times 2 \pi$.
Thus, the HST image only fails to test the holographic model by
3 orders of magnitude.

However, the absence of a quantum foam induced halo structure in the
HST image of PKS1413+135 rules out convincingly
the random-walk model.  (In fact, the scaling relation
discussed above indicates
that all spacetime foam models with $\alpha \lesssim 0.6$ are ruled
out by this HST observation.)  This result
has profound implications for cosmology. \cite{llo04,CNvD,Arzano}  To wit,
from the (observed) cosmic critical density in the present era,
a prediction of the holographic-foam-inspired cosmology,
we deduce that
$\rho \sim H_0^2/G \sim (R_H l_P)^{-2}$, 
where $H_0$ and $R_H$ are the present Hubble parameter and Hubble radius of
the observable universe respectively.
Treating the whole universe as a computer\cite{llo02, llo04}, one can
apply the Margolus-Levitin theorem to conclude that the universe
computes at a rate $\nu$ up to $\rho R_H^3 \sim R_H l_P^{-2}$
for a total of $(R_H/l_P)^2$
operations during its lifetime so far.
If all the information of this huge computer is stored in ordinary
matter, then we can apply standard methods of statistical mechanics
to find that the total number $I$ of bits is $(R_H^2/l_P^2)^{3/4} =
(R_H/l_P)^{3/2}$.
It follows that each bit flips once in the amount of time given by
$I/\nu \sim (R_H l_P)^{1/2}$.
On the other hand, the
average separation of neighboring bits is
$(R_H^3/I)^{1/3} \sim (R_H l_P)^{1/2}$.
Hence, 
the time
to communicate with neighboring bits is equal to the time for each
bit to flip once.  It follows that the accuracy to which ordinary
matter maps out the geometry of spacetime corresponds exactly to
the case of events
spread out uniformly in space and time discussed above for the case
of the random-walk model of quantum foam.  Succinctly,
ordinary matter only contains an amount of
information dense enough to map out spacetime at a level consistent with
the random-walk model. Observationally ruling out the random-walk model
suggests that there must be other kinds of matter/energy with which the
universe can map out its spacetime geometry to a finer spatial accuracy
than is possible with the use of ordinary matter.  This line of
reasoning then strongly hints at the existence 
of dark energy/matter independent of
the evidence from recent cosmological (supernovae, cosmic mircowave
background, gravitational lensing, galaxy configuration and 
clusters) observations.

Moreover, the fact that our universe is observed to be at or very close to
its critical energy density $\rho \sim (H/l_P)^2 \sim (R_H l_P)^{-2}$
must be taken as solid albeit indirect evidence
in favor of the holographic model because, as aforementioned, this model
is the only model that requires the energy density to be critical. The
holographic model also predicts
a huge number of degrees of freedom for the universe in the present era,
with the cosmic entropy given by \cite{Arzano} $I \sim H R_H^3/l_P^2 \sim
(R_H /l_P)^2$.
Hence the average energy carried by each bit is
$\rho R_H^3/I \sim R_H^{-1}$.  Such
long-wavelength bits or ``particles'' carry negligible kinetic energy.
Since pressure (energy density) is given by kinetic energy minus (plus)
potential energy, a negligible kinetic energy means that
the pressure of the unconventional energy is roughly equal to minus its
energy density, leading to accelerating cosmic expansion as has been
observed. This scenario is very similar to that
for quintessence.

How about the early universe?  Here
a cautionary remark is in order.
Recall that the holographic model has been derived for a static and flat
spacetime.  Its application to the universe of the present era may be valid,
but to extend the discussion to the early universe may need a judicious
generalization of some of the concepts involved.   
However, there is cause 
for optimism: for example,
one of the main
features of the holograpahic model, viz. the critical energy density, is
actually the hallmark of the inflationary universe paradigm.  Further 
study is warranted.

\section*{Acknowledgments}
This work was supported in part by the US Department of Energy
and the
Bahnson Fund of the University of North Carolina.

\end{document}